\documentclass[preprint,12pt]{elsarticle}
\biboptions{sort&compress}
\usepackage{amssymb}
\usepackage{enumitem}
\usepackage{bm}
\usepackage{lineno}

\def\KPIPIPI {\ensuremath{K^- \pi^+\pi^+\pi^-}}
\def\KPI {\ensuremath{K^-\pi^+}}
\def\KPIPIZ {\ensuremath{K^-\pi^+\pi^0}}
\def\KSPIPI{\ensuremath{K^0_{\rm S} \pi^+ \pi^-}}
\def\KSKPI{\ensuremath{K^0_{\rm S} K^\mp\pi^\pm}}

\journal{Physics Letters B}

\begin{document}

\begin{frontmatter}



\title{Improved determination of the $D \to K^-\pi^+\pi^+\pi^-$ coherence factor and associated hadronic parameters from a combination of $e^+e^-\to \psi(3770)\to c\bar{c}$ and $pp \to c \bar{c} X$ data}


%\author[madras]{J.~Libby\corref{cor1}}\cortext[cor1]{Corresponding author}
%\ead{libby@iitm.ac.in}
\author[oxford]{T.~Evans}
\author[bristol]{S.T.~Harnew}
\author[madras]{J.~Libby}
\author[oxford]{S.~Malde}
\author[bristol]{J.~Rademacker}
\author[oxford]{G.~Wilkinson\corref{cor1}}\cortext[cor1]{Corresponding author}
\ead{guy.wilkinson@physics.ox.ac.uk}

\address[oxford]{University of Oxford, Denys Wilkinson Building, Keble Road,  OX1 3RH, United Kingdom}
\address[bristol]{University of Bristol, Bristol, BS8 1TL, United Kingdom}
\address[madras]{Indian Institute of Technology Madras, Chennai 600036, India}
%\address[pnl]{Pacific Northwest National Laboratory, Richland, Washington 99352, USA}
%\address[wayne]{Wayne State University, Detroit, Michigan 48202, USA} 
%\address[cmu]{Carnegie Mellon University, Pittsburgh, Pennsylvania 15213, USA}
%\address[warwick]{University of Warwick, Coventry, CV4 7AL, United Kingdom}
%\address[luther]{Luther College, Decorah, Iowa 52101, USA}

\begin{abstract}
Measurements of the coherence factor  $R_{K3\pi}$, the average strong-phase difference $\delta^{K3\pi}_D$ and mean amplitude ratio $r_D^{K3\pi}$ for the decay  $D \to K^-\pi^+\pi^+\pi^-$ are presented.  These parameters are important inputs to the determination of the unitarity triangle angle $\gamma$ in $B^- \to DK^-$ decays, where $D$ designates a superposition of $D^0$ and $\bar{D}{}^0$ mesons decaying to a common final state. The results are based on a combined fit to observables obtained from a re-analysis of the CLEO-c $\psi(3770)$ data set and those  measured in a $D^0\bar{D}^0$ mixing study performed by the LHCb collaboration.
\end{abstract}


\end{frontmatter}

%\linenumbers

\section{Introduction}
\label{sec:intro}

Knowledge of the  $D \to K^{-}\pi^{+}\pi^{+}\pi^{-}$ coherence factor and associated hadronic parameters   is necessary for a measurement of the unitarity triangle angle $\gamma$  (also denoted $\phi_3$) in $B^- \to DK^-,  D \to K^{-}\pi^{+}\pi^{+}\pi^{-}$  decays. (The symbol $D$ is used to denote a neutral charm-meson that is not in a flavour eigenstate, or where the flavour eigenvalue is not relevant for the discussion.)

The coherence factor  $R_{K3\pi}$ and average strong-phase difference $\delta^{K3\pi}_D$ for the decay $D^{0}\to \KPIPIPI$ are defined as follows~\cite{ATWOODSONI}:
\begin{equation}
 R_{K3\pi}e^{-i\delta_{D}^{K3\pi}}  = 
 \frac{\int
\mathcal{A}^\ast_{K^{-}\pi^{+}\pi^+\pi^{-}}(\mathbf{x})\mathcal{A}_{K^{+}\pi^{-}\pi^+\pi^{-}}(\mathbf{x})\mathrm{d}\mathbf{x}}{A_{K^{-}\pi^{+}\pi^+
\pi^{-}}A_{K^{+}\pi^{-}\pi^{+}\pi^-}}\, ,
\end{equation}
where $\mathcal{A}_{K^{\pm}\pi^{\mp}\pi^{+}\pi^-}(\mathbf{x})$ is the decay amplitude of $D^{0}\to K^\pm\pi^\mp\pi^+\pi^-$ at a point in multi-body phase space described by parameters $\mathbf{x}$, and 
\begin{equation}
A_{K^{\pm}\pi^{\mp}\pi^{+}\pi^-}^{2}=\int |\mathcal{A}_{K^{\pm}\pi^{\mp}\pi^{+}\pi^-}(\mathbf{x})|^{2}\mathrm{d}\mathbf{x}.  
\end{equation}
Therefore $A_{K^{-}\pi^{+}\pi^{+}\pi^-}$ is the Cabibbo-favoured (CF) amplitude, averaged over phase space, and  $A_{K^{+}\pi^{-}\pi^{+}\pi^-}$ is the corresponding doubly Cabibbo-suppressed (DCS) quantity.
The coherence factor  takes a value between 0 and 1.  It is also useful to define the parameter $ r_{D}^{K3\pi} =  {A_{K^{+}\pi^{-}\pi^+\pi^{-}}}/{A_{K^{-}\pi^{+}\pi^{+}\pi^-}}$.  In this Letter the term `hadronic parameters' of the decay $D \to \KPIPIPI$ is employed to refer collectively to $R_{K3\pi}$, $\delta_D^{K3\pi}$ and $r_D^{K3\pi}$.    Throughout the discussion charge conjugation is implicit, the good approximation is made that $\mathit{CP}$-violation can be neglected in $D^0$ mixing and decay~\cite{HFAG}, and expressions are given in the convention $\mathit{CP}|D^0\rangle = |\bar{D}^0\rangle$. 

The role of the hadronic parameters in measurements sensitive to $\gamma$ can be appreciated by considering the decay  of $B^\mp$ mesons to a neutral charm-meson, reconstructed in the  inclusive $K^\pm\pi^\mp\pi^+\pi^-$ final state, and a charged kaon. 
Neglecting small   corrections from $D^0\bar{D}^0$ mixing, which can be included in a straightforward manner~\cite{RAMA}, the decay  rates are given by:
\begin{eqnarray}
{\Gamma(B^{\mp}\to(K^{\pm}\pi^{\mp}\pi^{+}\pi^-)_D K^{\mp})} &  \propto &
 (r_{B})^{2} + (r_{D}^{K3\pi})^{2}  +  \nonumber \\
& & 2r_{B}r_{D}^{K3\pi}R_{K3\pi}\cos{(\delta_{B}+\delta_D^{K3\pi}\mp \gamma)}\, .
\label{eqn:kpipi0adssuppressed} 
\end{eqnarray}
Here $r_B \sim 0.1$ is the absolute ratio of $B^- \to {\bar D}^0 K^-$  to $B^- \to D^0 K^-$ amplitudes.
The phase difference between these two amplitudes is $(\delta_B - \gamma)$, where $\delta_B$ is a $\mathit{CP}$-conserving strong phase.
The coherence factor, which is replaced by unity  in the equivalent expression for a single point in phase space or two-body $D$-meson decays, modulates the size of the interference term that carries the dependence on $\gamma$.

%The coherence factor and average strong-phase difference 
The hadronic parameters 
can be measured in the decays of coherently produced $D\bar{D}$ pairs at the $\psi(3770)$ resonance~\cite{ATWOODSONI}, such as are available in the data sets collected by the CLEO-c and BESIII experiments.  A double-tag technique is employed where one meson is reconstructed in the signal decay $D \to K^{-}\pi^{+}\pi^{+}\pi^{-}$, and the other in a tagging mode, for example, a $\mathit{CP}$ eigenstate.  This method was  applied by the CLEO collaboration to obtain first constraints on $R_{K3\pi}$ and $\delta^{K3\pi}_D$, together with  $R_{K\pi\pi^0}$ and $\delta^{K\pi\pi^0}_D$,  the analogous parameters for $D \to \KPIPIZ$ decays~\cite{WINGS}. A later analysis augmented the set of tags with the decay $D \to \KSPIPI$  in order to improve the sensitivity to the hadronic parameters of both signal modes~\cite{NEWWINGS}.  These results have been exploited by the LHCb collaboration to interpret a set of measurements with the decays $B^- \to DK^-$, $D \to \KPIPIPI$~\cite{LHCBK3PI} and $D \to \KPIPIZ$~\cite{LHCBNAZIM}, in combination with observables from other modes, to yield $\gamma = (73^{+9}_{-10})^\circ$~\cite{LHCBGAMMA}.

Observables sensitive to  $D^0\bar{D}^0$ mixing  in multibody $D$-meson decays  are also affected by the same hadronic parameters~\cite{SNEHA,SAM,SAM2}.  The time-dependent ratio  between $D^0 \to K^+\pi^-\pi^+\pi^-$ and $D^0 \to K^-\pi^+\pi^+\pi^-$ decay rates is
\begin{equation}
R(t) \approx (r_D^{K3\pi})^2 - a \frac{t}{\tau} + b \left(\frac{t}{\tau}\right)^2,
\label{eq:mixing}
\end{equation}
where $t$ is the proper decay time, $\tau$ is the mean $D^0$ lifetime, $a = R_{K3\pi}(y\cos\delta_D^{K3\pi}$ $- x\sin\delta_D^{K3\pi})$, $b = \frac{1}{4}(x^2 + y^2)$, and $x$ and $y$ are the commonly used mixing parameters, defined for example in Ref.~\cite{HFAG}.  As noted in Refs.~\cite{SAM,SAM2}, a measurement of $r_D^{K3\pi}$, $a$ and $b$ can either be used to determine $x$ and $y$, given external information on $R_{K3\pi}$ and $\delta_D^{K3\pi}$, or employed to set constraints on the latter parameters, provided $x$ and $y$ are known.   Both interpretations are presented in Ref.~\cite{LHCBMIX}, which reports a first observation of $D^0\bar{D}^0$ mixing with this multibody final state.
  
Improved knowledge of $R_{K3\pi}$,  $\delta^{K3\pi}_D$ and $r_D^{K3\pi}$ is also needed to help understand the contribution of  $D \to K^{-}\pi^{+}\pi^{+}\pi^{-}$ decays to the width difference in the $D^0\bar{D}^0$ system~\cite{GLW}.

This Letter reports  a new analysis of $D \to \KPIPIPI$ decays in the CLEO-c data set,  and benefits from an updated Monte Carlo simulation sample to correct a biased estimate of the background contamination in several sets of double tags that afflicted the results reported in Refs.~\cite{WINGS,NEWWINGS}.  In addition, the quasi-$\mathit{CP}$-eigenstate $D\to\pi^+\pi^-\pi^0$~\cite{FPLUSPPP0} is included as a new tagging mode to augment the sensitivity of the analysis. 
Furthermore, an error is corrected in the formalism used to interpret the yields found with the $D \to \KSPIPI$ tags~\footnote{The error is present in Ref.~\cite{NEWWINGS} and also in the previously posted version of this paper.}.
The observables from this new analysis are then used to determine updated constraints on the hadronic parameters.   Finally, a global fit is made to the observables from this study and those measured in $D^0\bar{D}^0$ mixing by LHCb, to obtain more precise results for the $D \to \KPIPIPI$ hadronic parameters.  
Since one of the re-measured $\psi(3770)$ observables couples the \KPIPIPI\ and \KPIPIZ\ systems,  it is also possible to determine updated values of   $R_{K\pi\pi^0}$, $\delta_D^{K\pi\pi^0}$ and $r_D^{K\pi\pi^0}$.

\section{Measuring the hadronic parameters with CLEO-c data}

In this section the observables sensitive to the hadronic parameters are reviewed, and a new analysis of the CLEO-c data set is presented that updates the measured values with respect to those reported in Ref.~\cite{NEWWINGS}.  Information on the hadronic parameters is then obtained from a fit to the updated measurements.

\subsection{Observables}

Two categories of observables exist, both based on double-tagged measurements in which one $D$ meson decays to the signal mode \KPIPIPI, and the other decays to a tagging mode.  In the first category the tagging modes are two-body or higher multiplicity final states, which are treated in an inclusive fashion.  In the second case, the tagging mode is the self-conjugate final-state \KSPIPI\ and yields are measured in different  phase-space bins of this tagging decay, defined in the plane of the Dalitz plot.

In the category involving inclusive tags, so-called $\rho$ observables are constructed, which are the ratio of the measured  double-tag yields, after background subtraction and efficiency correction, to the yields expected in the absence of quantum-correlations.  
\begin{description}[style=nextline]
\item[$\bm { \rho^{K3\pi}_{\mathit{CP}\pm}}$] {These are the  ratios where the tagging mode is a $\mathit{CP}$-even eigenstate ($\rho^{K3\pi}_{\mathit{CP}+}$) or a $\mathit{CP}$-odd eigenstate ($\rho^{K3\pi}_{\mathit{CP}-}$).  Neglecting  corrections from $D^0\bar{D}^0$ mixing, which enter through the definitions of the branching fractions used in the normalisation factors,
$\rho^{K3\pi}_{\mathit{CP}\pm} = 1  \mp 2 r_D^{K3\pi}R_{K3\pi} \cos \delta_D^{K3\pi}/$ $[1 + (r_D^{K3\pi})^2].$  Precise definitions and mixing-corrected expressions for this, and subsequent relations, can be found in Ref.~\cite{WINGS}.

} 
\item[$\bm { \Delta^{K3\pi}_{\mathit{CP}}}$]{
This is a $\mathit{CP}$-invariant observable defined $ \Delta^{K3\pi}_{\mathit{CP}} \equiv \pm 1  \times (\rho^{K3\pi}_{\mathit{CP} \pm} - 1) .$ It allows  the results for the $\mathit{CP}$-even and $\mathit{CP}$-odd tags to be combined together. Neglecting mixing, $\Delta^{K3\pi}_{\mathit{CP}} = -2 r_D^{K3\pi} R_{K3\pi} \cos  \delta_D^{K3\pi}/[1 + (r_D^{K3\pi})^2]$. }
\item[$\bm { \rho^{K3\pi}_{LS}}$]{  This is the ratio involving \KPIPIPI\ vs.  \KPIPIPI\  events, {\it i.e.} those where the kaons are of like sign (LS).
 Neglecting mixing,  $\rho^{K3\pi}_{LS} = 1 - (R_{K3\pi})^2.$
 }
\item[$\bm { \rho^{K3\pi}_{K\pi, LS}}$]{ This is the ratio involving \KPIPIPI\ vs. \KPI\  events. Neglecting mixing,
$ \rho^{K3\pi}_{K\pi, LS} = 1 \,- \, 2 (r_D^{K3\pi}/r_D^{K\pi}) R_{K3\pi} \cos(\delta_D^{K\pi} - \delta_D^{K3\pi})/\left[1 + (r_D^{K3\pi}/r_D^{K\pi})^2 \right]$, where $r_D^{K\pi}$ is the ratio between the  DCS and CF  $D  \to K^-\pi^+$ amplitudes and  $\delta^{K\pi}_D$  the accompanying strong-phase difference.}

\item[$\bm { \rho_{K\pi\pi^0,LS}^{K3\pi}}$]{  This is the ratio involving \KPIPIPI\ vs. \KPIPIZ\  events. Neglecting mixing, 
$\rho_{K\pi\pi^0,LS}^{K3\pi} = 1 \,- \, 2 (r_D^{K3\pi}/r_D^{K\pi\pi^0}) R_{K3\pi} R_{K\pi\pi^0} \cos(\delta_D^{K\pi\pi^0} - \delta_D^{K3\pi})/$ $[1 + (r_D^{K3\pi}/r_D^{K\pi\pi^0})^2 ].$}
\end{description}
 The deviation of any of the $\rho$ observables from unity or $\Delta^{K3\pi}_{\mathit{CP}}$ from zero  is indicative of a non-zero coherence factor.   Recalling that $r_D^{K3\pi}$, $r_D^{K\pi}$ and $r_D^{K\pi\pi^0} \sim 0.05$, it is expected that $|\Delta^{K3\pi}_{\mathit{CP}}| < 0.1$, whereas larger effects can occur for the like-sign observables.

The observables involving  \KSPIPI\ decays  comprise the yields of double tags, after background subtraction and bin-to-bin relative efficiency correction, in eight pairs of bins in the plane of the Dalitz plot symmetric about the line $m[K^0_{\rm S} \pi^+]^2 =  m[K^0_{\rm S} \pi^-]^2$.  The binning scheme follows the   `equal $\Delta \delta_D$' definition of Ref.~\cite{CLEOKSPIPI,BONDAR}, where the partitioning is guided by an amplitude model developed by the BaBar collaboration~\cite{BABAR_2008}.  The observables are denoted $Y_i$, where the subscript gives the bin number ($i=-8 \to 8$, excluding 0).
The values of  $Y_i$  differ from those expected in the incoherent case in a manner that is dependent on the values of the coherence factor and average strong-phase difference of the signal mode~\cite{NEWWINGS}:
\begin{eqnarray}
Y_i & = & H_{K3\pi} {\Big ( }  K_i + (r_D^{K3\pi})^2K_{-i} -  \nonumber \\
& & 2 r_D^{K3\pi} \sqrt{K_i K_{-i}} R_{K 3\pi} [c_i \cos \delta_D^{K3\pi} - s_i \sin \delta_D^{K3\pi}] {\Big )}.
\label{eq:kspipiyield}
\end{eqnarray}
(In Ref.~\cite{NEWWINGS} this expression is written with an incorrect sign before the term $ s_i \sin \delta_D^{K3\pi}$).
Here $H_{K3\pi}$ is a bin-independent normalisation factor and $K_i$ is the fractional yield of $D^0$ decays that fall into bin $i$. The parameters $c_i$ and $s_i$ are  the amplitude-weighted averages over bin $i$ of $\cos(\Delta \delta_D)$ and $\sin(\Delta \delta_D)$, respectively, where $\Delta \delta_D$ is the strong-phase difference between the $D^0 \to \KSPIPI$ and $\bar{D}^0 \to  \KSPIPI$ amplitudes at a single point in the  Dalitz plot. All these $D \to K^0_{\rm S} \pi^+\pi^-$ quantities are defined ignoring $D^0\bar{D}{}^0$ mixing effects, which is appropriate for double-tag yields arising from $\psi(3770)$ mesons produced at rest in the laboratory, as is the case for the CLEO-c experiment~\cite{ANTONMIX}.

\subsection{Yield determination and results for observables}

An 818~$\rm pb^{-1}$ data set of $e^+e^-$ collisions produced by the Cornell Electron Storage Ring (CESR) at $\sqrt{s}=3.77$~GeV and collected with the CLEO-c detector is analysed.   The CLEO-c detector is described in detail elsewhere~\cite{CLEOC}.  In addition, simulated Monte Carlo samples are studied to assess possible background contributions and to determine efficiencies. The EVTGEN package~\cite{EVTGEN} is used to generate the decays and GEANT~\cite{GEANT} is used to simulate the CLEO-c detector response.

To ensure full understanding of all the inputs to the analysis, the selection of all double tags involving the decay $D\to \KPIPIPI$ is re-performed, although the selection criteria are intended to be identical to those reported in Refs.~\cite{WINGS,NEWWINGS}.    The full list of final states reconstructed is given in Table~\ref{tab:tags}, with $\pi^0 \to \gamma \gamma$, $K^0_{\rm S} \to \pi^+\pi^-$, $\phi \to K^+K^-$, $\eta \to \gamma\gamma$, $\eta \to \pi^+\pi^-\pi^0$ and $\eta^\prime \to \eta (\gamma\gamma) \pi^+\pi^-$.  
A single new addition to the list of tag modes is the abundant decay $D \to \pi^+\pi^-\pi^0$, which has recently been measured to be very close to a $\mathit{CP}$ eigenstate, with a $\mathit{CP}$-even fraction of $F_+^{\pi\pi\pi^0} = 0.973 \pm 0.017$~\cite{FPLUSPPP0}.  The selection requirements for this new mode are similar to those reported in Ref.~\cite{FPLUSPPP0}.  In particular, a $K^0_{\rm S}$ veto is applied to the $\pi^+\pi^-$ combination, as described in Ref.~\cite{WINGS}, in order to suppress contamination from $D \to K^0_{\rm S} \pi^0$ decays.  This veto rejects candidates where the two-track vertex is significantly displaced from the beamspot, or either track has a significant impact parameter.    

\begin{table}[htb]
\caption{$D$-meson final states reconstructed in the analysis.}\label{tab:tags}
\begin{center}
\begin{tabular}{cc} \hline\hline
Type  & Final states \\ \hline
\noalign{\vskip 0.075cm}
Flavoured & \KPI\, \KPIPIPI\, \KPIPIZ \\
$\mathit{CP}$ even & $K^+K^-$, $\pi^+\pi^-$, $K^0_{\rm S} \pi^0\pi^0$, $K_{\rm L}^0 \pi^0$, $K_{\rm L}^0 \omega$,  $\pi^+\pi^-\pi^0$ \\
$\mathit{CP}$ odd   & $K^0_{\rm S} \pi^0$, $K^0_{\rm S} \omega$, $K^0_{\rm S} \phi$, $K^0_{\rm S} \eta$, $K^0_{\rm S} \eta^\prime$ \\
Self conjugate & $K^0_{\rm S} \pi^+\pi^-$ \\ \hline
\end{tabular}
\end{center}
\end{table}

\begin{figure}[htb]
\begin{center}
  \includegraphics[width=0.48\columnwidth]{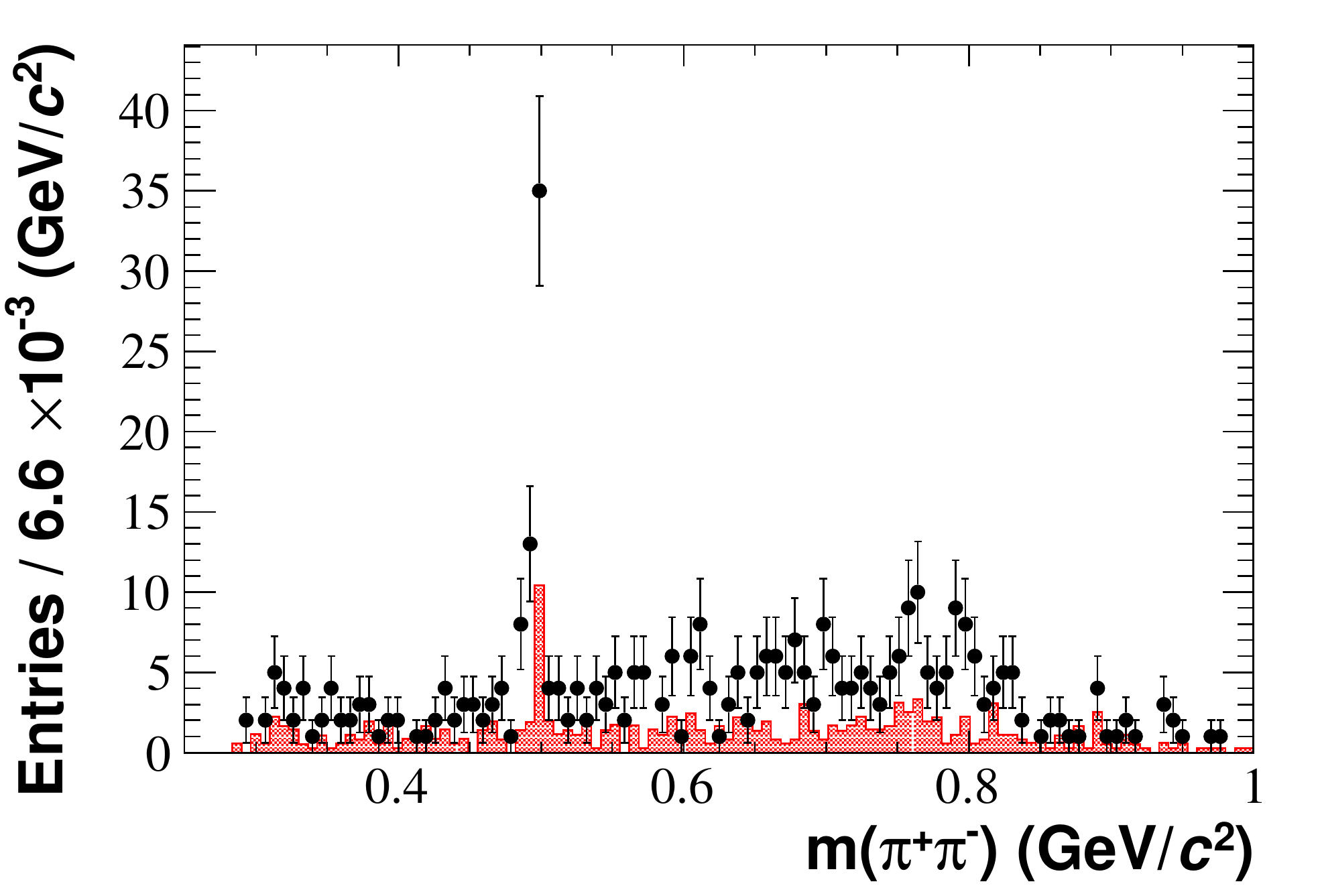}
  \includegraphics[width=0.48\columnwidth]{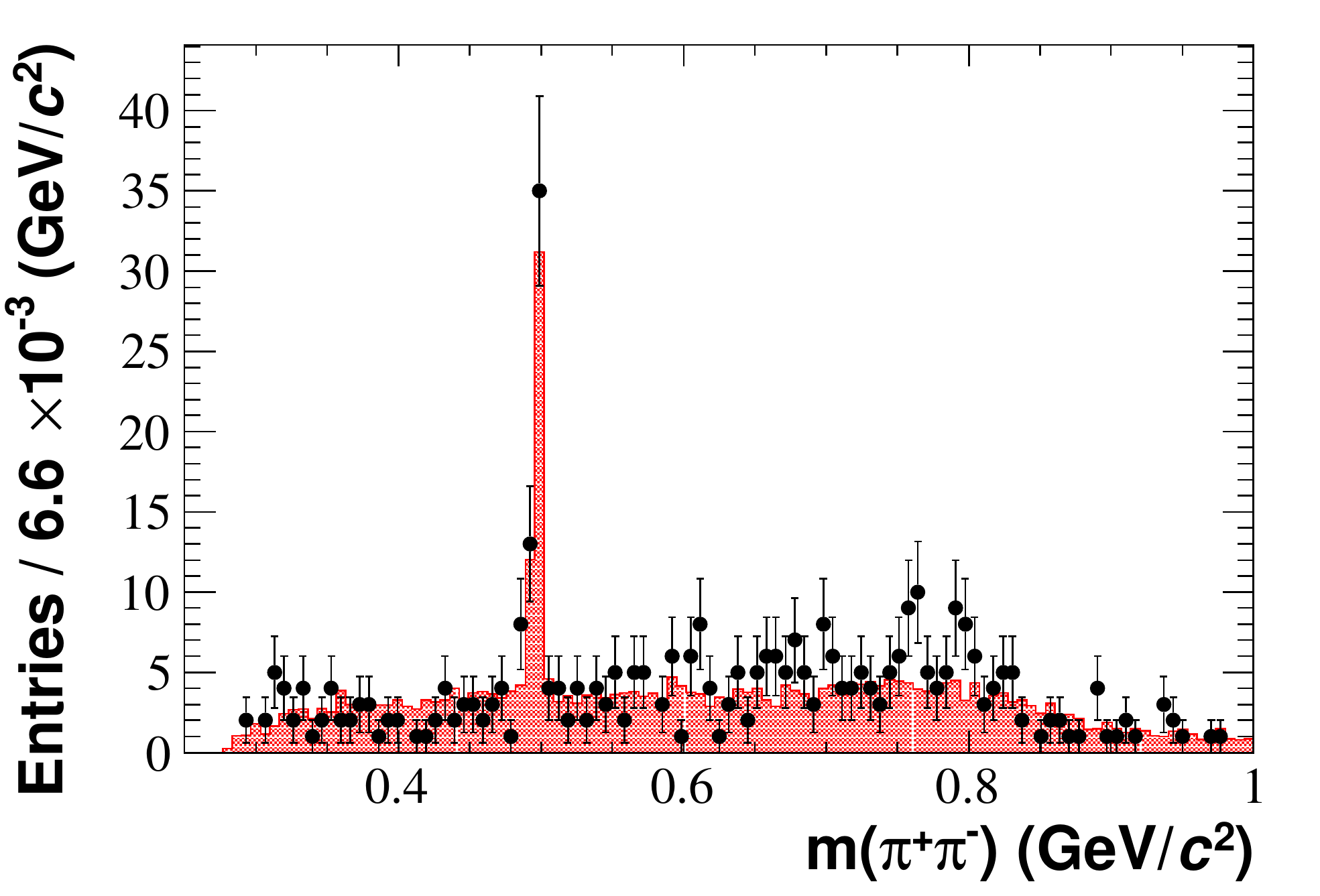}
\end{center}
\caption{Invariant mass of the $\pi^+\pi^-$ pairs from the $D \to \KPIPIPI$ candidates summed over the \KPIPIPI\ vs. \KPIPIPI, \KPIPIPI\ vs. \KPI\ and \KPIPIPI\  vs. \KPIPIZ\ double tags.   Data are shown by points, and the simulation by the filled histogram.  Left: old simulation used in Ref.~\cite{WINGS}.  Right: new simulation. }
\label{fig:pipimasses}
\end{figure}

The most significant change in the analysis concerns the sample of simulated inclusive $D^0\bar{D}^0$ events used to estimate the contamination from specific background decays that occur in or close to the kinematic region where the signal peaks.  The sample  in the new analysis  is a factor of two larger than that used in the original studies and benefits from updated knowledge of branching fractions.   
The  singly Cabibbo-suppressed modes $D \to K^0_{\rm S} K^\mp \pi^\pm$ (here specifying explicitly the two final states) are a dangerous  source of background for the like-sign $ \rho^{K3\pi}_{LS}$, $ \rho^{K3\pi}_{K\pi, LS}$
  and $ \rho^{K3\pi}_{K\pi\pi^0, LS}$ observables since $\frac{{\cal B}(D^0 \to K^0_{\rm S} K^+ \pi^- )}{{\cal B} (D^0 \to K^0_{\rm S} K^- \pi^+ )} \sim {\cal O}(1)$, whereas $\frac{{\cal B} (D^0 \to K^+\pi^-\pi^-\pi^+)}{{\cal B} (D^0 \to K^- \pi^+\pi^+\pi^-)}  \sim {\cal O}(10^{-3})$~\cite{PDG}.
These modes were incorrectly simulated in the old Monte Carlo  sample, being generated at a rate that was a factor of three lower than the measured branching fractions~\cite{PDG}, and  with a resonant substructure that poorly matches experimental results~\cite{CLEOCKSKPI,LHCBKSKPI}.
Both of these deficiencies are corrected in the new simulation.  Figure~\ref{fig:pipimasses} shows the invariant mass of the $\pi^+\pi^-$ combinations, summed over all like-sign double tags, for both data and simulation.  
The selection requirements include a $K^0_{\rm S}$ veto on the $\pi^+\pi^-$ combination in the signal-decay candidate to suppress $D \to \KSKPI$ contamination.
A clear peak is seen from the residual background surviving the $K^0_{\rm S}$ veto, which is well modelled by the new Monte Carlo sample, but was previously  described poorly.  (The old Monte Carlo simulation also contained other deficiencies, apparent from poor agreement in other regions of the $\pi^+\pi^-$ mass spectrum, but these did not impact directly upon the analysis.)

The event yields after background subtraction  are presented in Table~\ref{tab:inclusive}. When appropriate, the contamination from peaking background is corrected for the small effects of quantum-correlation, which are not simulated  in the Monte Carlo.   The contribution of non-peaking background is determined in a data-driven manner, as in the original analysis~\cite{WINGS}.   The yields for \KPIPIPI\ vs. \KPIPIPI, \KPIPIPI\ vs. \KPI\ and \KPIPIPI\  vs. \KPIPIZ\ are all significantly lower than previously reported, because of the revised estimate for the level of $D \to \KSKPI$ contamination, which is the dominant source of background for these double tags, and is now determined to comprise around 36\%, 30\% and 30\%, respectively,  of the selected events in each of the three samples.  
For all the other classes of double tags, where the mean purity is in excess of 95\%, the differences in results with respect to the earlier analysis are either negligible, or small and well understood. 

\begin{table}[htb]
\caption{Measured double-tagged yields and statistical uncertainties after background subtraction.}
\label{tab:inclusive}
\begin{center}
\begin{tabular}{l  cc}
\hline\hline
Mode & \KPIPIPI\ & \KPI\ \\
\hline
$K^+\pi^-\pi^+\pi^-$ &4006.3  $\pm$  65.0   & \phantom{xx}-- \\
$K^-\pi^+\pi^+\pi^-$ & \phantom{x}19.7  $\pm$   6.2   &  \phantom{xx}-- \\
$K^+\pi^-$                  &5203.7  $\pm$   72.7  & 1723.1 $\pm$ 41.8  \\
$K^-\pi^+$                  &\phantom{x}26.6  $\pm$   6.2 &\phantom{xx}--  \\
$K^+\pi^-\pi^0$         &10598.0  $\pm$  104.8  & \phantom{xx}--  \\
$K^-\pi^+\pi^0$         &\phantom{x}53.1  $\pm$  9.1 & \phantom{xx}-- \\ \hline
%& & \\
$K^+K^-$                    &\phantom{x}542.0  $\pm$  23.4  & \phantom{xx}-- \\
$\pi^+\pi^-$                &\phantom{x}244.2  $\pm$ 15.9 & \phantom{xx}--\\
$K^0_{\rm S}\pi^0\pi^0$  &\phantom{x}299.5  $\pm$   18.3 & \phantom{x}223.5 $\pm$ 15.5  \\
$K^0_{\rm L}\pi^0$        &  \phantom{x}839.4  $\pm$  30.6 &  \phantom{x}703.0 $\pm$ 27.9\\
$K^0_{\rm L}\omega$  & \phantom{x}302.8    $\pm$  19.0 &  \phantom{x}247.3 $\pm$ 17.0\\
$\pi^+\pi^-\pi^0$              &1280.0  $\pm$  37.2 & \phantom{x}951.9 $\pm$ 31.4\\ \hline
%&&\\
$K^0_{\rm S}\pi^0$     &\phantom{x}701.3  $\pm$   26.9  &\phantom{x}472.5 $\pm$ 21.8 \\
$K^0_{\rm S}\omega$  & \phantom{x}340.7  $\pm$   19.8  &\phantom{x}202.0 $\pm$ 15.3\\
$K^0_{\rm S}\phi$            & \phantom{x}57.5  $\pm$  8.0  &\phantom{x}47.8 $\pm$ 7.3\\
$K^0_{\rm S}\eta(\gamma \gamma)$ & \phantom{x}135.0  $\pm$  12.1  &\phantom{x}67.2 $\pm$ 8.4  \\
$K^0_{\rm S}\eta(\pi^+\pi^-\pi^0)$  & \phantom{x}37.5  $\pm$  7.2  &  \phantom{x}27.2 $\pm$ 5.8\\
$K^0_{\rm S}\eta^\prime(\pi^+\pi^-\eta)$ & \phantom{x}40.1   $\pm$  6.4  & \phantom{x}31.7 $\pm$ 5.7 \\ \hline
$K^0_{\rm S} \pi^+\pi^-$ & 2206.4 $\pm$ 48.6 & \phantom{xx}-- \\
\hline\hline
\end{tabular}
\end{center}
\end{table}

The normalisation for the like-sign observables is performed through measurement of the corresponding opposite-sign yields ({\it e.g.} $K^- \pi^+ \pi^+\pi^-$ vs. $K^+ \pi^- \pi^+\pi^-$), which are negligibly modified by quantum-correlation effects, and from knowledge of the ratios of the relevant charm-meson branching fractions.  The normalisation for the $\mathit{CP}$ observables $\rho^{K3\pi}_{\mathit{CP}\pm}$ is performed  in one of two ways.  For the $D \to K^+K^-$ and $D \to \pi^+\pi^-$ tags the expected number of events in the incoherent limit is calculated through knowledge of the branching ratios, the total number of $D\bar{D}$ events, as determined through the yield of opposite sign double-tags, and the reconstruction efficiency, as measured from simulation.  For the other tags, where the branching fractions and reconstruction efficiencies are less well known, the normalisation makes use of the corresponding number of events where the $\mathit{CP}$-tag is reconstructed together with a $D \to \KPI$ decay.  This treatment requires corrections for quantum-correlation effects in the normalisation mode, which introduces minor dependence on the $D^0\bar{D}^0$ mixing parameters and the ratio $r_D^{K\pi}$ between the  DCS and CF  $D  \to K^-\pi^+$ amplitudes and accompanying strong-phase difference $\delta^{K\pi}_D$.  Again, full details can be found in Ref.~\cite{WINGS}.   The external values used in these determinations, and their sources, are  summarised in Table~\ref{tab:inputs}.   There are several small updates with respect to the values used in Ref.~\cite{NEWWINGS}, none of which induce significant changes on the results.

\begin{table}[ht]
 \caption{Values of branching fractions and other parameters  used in the determination of the CLEO-c observables.}\label{tab:inputs}
\begin{center}
\begin{tabular}{lcc} \hline\hline
Parameter & Value & Reference \\ \hline
\noalign{\vskip 0.075cm}
$\mathcal{B}(D^{0}\to K^{-}\pi^{+}\pi^{+}\pi^{-})$ & $(8.29\pm 0.20)\%$ & \cite{NEWCLEO}  \\
$\frac{\mathcal{B}(D^{0}\to K^{+}\pi^{-}\pi^{-}\pi^{+})}{\mathcal{B}(D^{0}\to K^{-}\pi^{+}\pi^{+}\pi^{-})}$ & $(3.25\pm 0.11)\times 10^{-3}$ & \cite{PDG} \\
%&&\\
%$\mathcal{B}(D^{0}\to K^{-}\pi^{+}\pi^{0})$ & $(14.96\pm 0.34)\%$ & \cite{NEWCLEO}  \\
$\frac{ \mathcal{B}(D^{0}\to K^{+}\pi^{-}\pi^{0})} {\mathcal{B}(D^{0}\to K^{-}\pi^{+}\pi^{0}) }$ & $(2.20\pm 0.10)\times 10^{-3}$ & \cite{PDG}  \\
%&& \\
%%$\mathcal{B}(D^{0}\to K^{-}\pi^{+})$ & $(3.88\pm 0.05)\%$ & \cite{PDG}  \\
$(r_{D}^{K\pi})^2$ & $(0.349 \pm 0.004) \%$ & \cite{HFAG}  \\ 
 $\delta_{D}^{K\pi}$ & $(191.8^{+\phantom{0}9.5}_{-14.7})^{\circ}$ & \cite{HFAG} \\ 
%& & \\
 $x$ & $(0.37\pm 0.16)\%$ & \cite{HFAG} \\
 $y$ & $(0.66^{+0.07}_{-0.10})\%$ & \cite{HFAG} \\
%& & \\
$\mathcal{B}(D^{0}\to K^+K^-)$ & $(3.96\pm 0.08) \times 10^{-3}$ & \cite{PDG}  \\
$\mathcal{B}(D^{0}\to \pi^+\pi^-)$ & $(1.402\pm 0.026) \times 10^{-3}$ & \cite{PDG}  \\
%& & \\
$F^{\pi\pi\pi^0}_+$ & $0.973 \pm 0.017$ & \cite{FPLUSPPP0} \\
%$F^{K^0_{\rm S} \phi}_+$ & $< 0.33$ & \cite{CLEOKSPIPI} \\
\hline\hline
\end{tabular}
\end{center}
\end{table}

The assignment of systematic uncertainties for the inclusive tags follows the same  procedure as applied  in Ref.~\cite{WINGS}, where contributions arise from uncertainties in the external parameters, the finite size of the samples used in the various normalisations,  knowledge of reconstruction efficiencies (relevant only for the double tags involving the modes $D \to K^+K^-$ and $D \to \pi^+\pi^-$), assumptions involved in the $D \to K^-\pi^+$ normalisation procedure, and the potential bias from non-uniform acceptance across the phase space of the signal mode.  
For $\rho^{K3\pi}_{\mathit{CP}\pm}$ and $\Delta_{\mathit{CP}}^{K3\pi}$ the leading source of systematic error comes from the uncertainties in the yields of the $D \to K^-\pi^+$ normalisation samples.
An important new component, dominant for the like-sign observables, accounts for a $\pm 20\%$ uncertainty in the level of residual contamination from $D \to \KSKPI$ decays.  In addition, small new contributions are assigned associated with the finite knowledge of the $\mathit{CP}$-impurity in $D \to \pi^+\pi^-\pi^0$ and the potential non-$\phi$ contribution to the $D  \to K^0_{\rm S} K^+K^-$ sample~\cite{CLEOKSPIPI}, and incomplete understanding of the effects of quantum-correlation on the background sources.

\begin{figure}[htb]
\begin{center}
\vspace*{-1.5cm}
  \includegraphics[width=0.7\columnwidth]{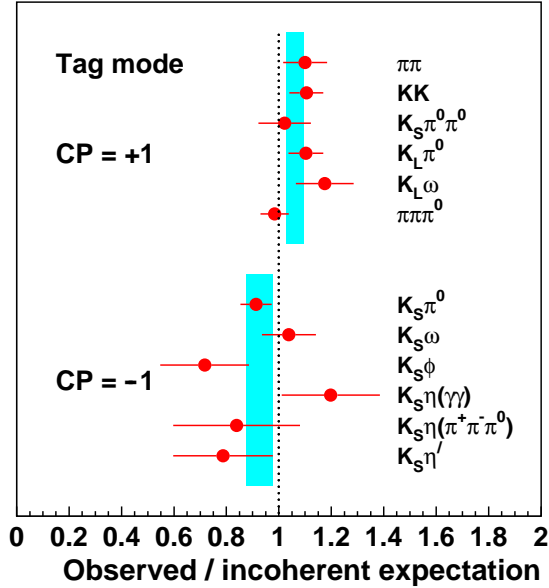}
\vspace*{-3.0cm}
\end{center}
\caption{Results for the $\rho^{K3\pi}_{\mathit{CP}+}$ and  $\rho^{K3\pi}_{\mathit{CP}-}$ observables for each tag.  The error bars give the total uncertainty on the individual measurements.  The blue bands represent the 1~$\sigma$ bound on the averaged results for each observable.}
\label{fig:cptagresults}
\end{figure}

Separate values of $\rho^{K3\pi}_{\mathit{CP}\pm}$ and $\Delta^{K3\pi}_{\mathit{CP}}$ are calculated for each $\mathit{CP}$ tag.   In the case of $\pi^+\pi^-\pi^0$ a correction factor of $(2F^{\pi\pi\pi^0}_+ -1 )^{-1}$ is applied to the raw result for $\Delta^{K3\pi}_{\mathit{CP}}$  to account for residual $\mathit{CP}$-odd contributions to this tagging mode.  
The individual results for $\rho^{K3\pi}_{\mathit{CP}+}$ and  $\rho^{K3\pi}_{\mathit{CP}-}$ are displayed in Fig.~\ref{fig:cptagresults}.  The results for the mean value of these quantities, and for the $\mathit{CP}$-invariant observable $\Delta^{K3\pi}_{\mathit{CP}}$, evaluated taking full account of correlations are given in Table~\ref{tab:observables}.  The $\chi^2$ for the twelve measurements in the $\Delta^{K3\pi}_{\mathit{CP}}$ combination is 10.3, which indicates good compatibility.  The result for  $\Delta^{K3\pi}_{\mathit{CP}}$ is around $1 \sigma$ lower than formerly.  This shift can almost wholly be  attributed to the inclusion of the $D \to \pi^+\pi^-\pi^0$ tag, which returns a value for $\rho^{K3\pi}_{\mathit{CP}+}$ lower than that of the other $\mathit{CP}$-even tags, although still compatible.  If this contribution is excluded then the average result becomes  $\Delta^{K3\pi}_{\mathit{CP}} = 0.087 \pm 0.018 \pm 0.023$, which is in excellent agreement with that found in Ref.~\cite{NEWWINGS}.

All the values of the like-sign observables, also presented in Table~\ref{tab:observables}, are around 1.5$\sigma$ lower than before.  These shifts are a consequence of the improved understanding of the background involving $D \to \KSKPI$ decays.

\begin{table}[thb]
\caption{Measured values of the $\rho$ and $\Delta$ observables, as determined in the current analysis and reported in the previous analysis~\cite{NEWWINGS}. Here the first uncertainty is statistical and the second systematic.}
\label{tab:observables}
\begin{center}
\begin{tabular}{l  c c c} \hline\hline
Observable & Measured value & \hspace*{0.1cm} &    Previous result \\ \hline
\noalign{\vskip 0.075cm}
$\rho^{K3\pi}_{\mathit{CP}+}$         & $1.061$    $\pm$  $0.019$   $\pm$  $0.028$  &  &  $1.087$  $\pm$  $0.024$  $\pm$  $0.029$ \\
$\rho^{K3\pi}_{\mathit{CP}-}$         & $0.926$   $\pm$  $0.027$   $\pm$  $0.042$   && $0.934$    $\pm$  $0.027$  $\pm$  $0. 046$ \\
$\Delta^{K3\pi}_{\mathit{CP}}$ &  $0.063$         $\pm$   $0.015$   $\pm$  $0.021$  & &  $0.084$       $\pm$   $0.018$  $\pm$  $0.022$  \\
$\rho^{K3\pi}_{LS}$          & $0.757$     $\pm$  $0.239$   $\pm$  $0.122$ && $1.116$     $\pm$  $0.227$  $\pm$  $0.073$ \\ 
$\rho^{K3\pi}_{K\pi, LS}$    & $0.719$   $\pm$  $0.168$  $\pm$  $0.077$ && $1.018$   $\pm$  $0.177$  $\pm$  $0.054$ \\
$\rho_{K\pi\pi^0,LS}^{K3\pi}$&    $0.919$  $\pm$   $0.158$   $\pm$  $0.098$ & &    $1.218$  $\pm$   $0.169$   $\pm$  $0.062$   \\ \hline \hline
\end{tabular}
\end{center}
\end{table}

The correlation matrix for the like-sign and $\Delta^{K3\pi}_{\mathit{CP}}$ observables and for those of the analogous quantities for the decay $D \to \KPIPIZ$ may be found in~\ref{sec:cleoccorrel}.

Each double-tagged event involving $D \to K^0_{\rm S} \pi^+\pi^-$ decays is subjected to a mass-constrained fit of the tagging candidate in order to determine more reliably its location in the Dalitz plot, and hence the bin assignment. The background contamination in this sample is below 10\% in all bins. A large sample of Monte Carlo signal events is used to determine the relative bin-to-bin efficiencies, which all differ by less than 5\%.  The resulting values of the $Y_i$ observables, after background subtraction and relative efficiency correction, are presented in Table~\ref{tab:k0pipi_cor}.  All sources of systematic bias are negligible compared with the statistical uncertainties.

\begin{table}[tbh]
\caption{The $K^0_{\rm S}\pi^+\pi^-$-tagged signal yields, corrected for relative bin-to-bin efficiency effects, normalised to the  bin of highest efficiency.}\label{tab:k0pipi_cor}
\begin{center}
\begin{tabular}{cccc} \hline\hline
 Bin & $Y_i$ & Bin &  $Y_i$  \\ \hline 
1 & 354.4 $\pm$ 21.3 & $-$1 & 180.4 $\pm$ 15.0 \\
2 &  217.2 $\pm$ 16.1 & $-$2 &  56.8 $\pm$ 8.2  \\
3 &  183.8 $\pm$ 13.7 & $-$3 & 45.8 $\pm$   7.0  \\
4 &    62.1 $\pm$  8.3   & $-$4 & 41.7 $\pm$   6.8  \\
5 &  179.2 $\pm$ 14.5 & $-$5 &\phantom{x}96.9 $\pm$  10.6  \\
6 &  110.4 $\pm$ 11.3 & $-$6 & 33.7 $\pm$  6.3  \\
7 &  290.8 $\pm$ 18.9 & $-$7 & 35.7 $\pm$   6.5  \\
8 & 293.5 $\pm$ 19.1 & $-$8 & 75.9 $\pm$  9.4  \\ \hline\hline
\end{tabular}
\end{center}
\end{table}

\subsection{Fit to the coherence factor and average strong-phase difference}
\label{sec:cleofit}

The measured values of the $D \to \KPIPIPI$ observables, reported in Tables~\ref{tab:observables} and~\ref{tab:k0pipi_cor}, are input to a $\chi^2$ fit to determine $R_{K3\pi}$, $\delta_{D}^{K3\pi}$ and $r_D^{K3\pi}$.   The observable  $\rho_{K\pi\pi^0,LS}^{K3\pi}$ couples the $\KPIPIPI$ results to those of the $D \to \KPIPIZ$ system. Therefore the observables specific to  the latter decay, with values taken from Ref.~\cite{NEWWINGS}, are also included in the fit, and $R_{K\pi\pi^{0}}$, $\delta_{D}^{K\pi\pi^{0}}$ and $r_D^{K\pi\pi^0}$ treated as free parameters.~\footnote{In previous studies~\cite{WINGS,NEWWINGS} $r_D^{K3\pi}$ and $r_D^{K\pi\pi^0}$ were not fit parameters.  The change of strategy is motivated by the importance of these parameters in the $\gamma$ determination, and by the additional information made available through the LHCb $D \to \KPIPIPI$  mixing analysis~\cite{LHCBMIX}.} All known sources of correlation are accounted for.  

Full expressions relating the inclusive observables to the underlying physics parameters can be found in Ref.~\cite{WINGS}, and involve not only the hadronic parameters of the two multi-body signal modes, but also $r_D^{K\pi}$, $\delta_D^{K\pi}$, $x$ and $y$.  
These four latter parameters are therefore also floated in the fit, but with Gaussian constraints to the values of external measurements.
The ratio of the `wrong sign' to the `right sign' decay-time integrated branching fractions, $\frac{\mathcal{B}(D^{0}\to K^{+}\pi^{-}\pi^{-}\pi^{+})}{\mathcal{B}(D^{0}\to K^{-}\pi^{+}\pi^{+}\pi^{-})}$,  is an important additional measurement.  This observable can be related to  $r_D^{K3\pi}$ and the other two hadronic parameters, together with $x$ and $y$, through the time-integrated form of Eq.~\ref{eq:mixing}, as given in Ref.~\cite{NEWWINGS}  (and analogously for the $D^0 \to \KPIPIZ$ quantities).   The values of the external measurements taken for the ratios of the branching fractions, and for the Gaussian constraints of the additional fit parameters, are listed in Table~\ref{tab:inputs}.

Equation~\ref{eq:kspipiyield} is used to interpret the $Y_i$ observables in terms of  $R_{K3\pi}$, $\delta_{D}^{K3\pi}$ and $r_D^{K3\pi}$ (and an analogous  expression is employed for the \KPIPIZ\ case). 
Here also there are additional parameters, all associated with the $K^0_{\rm S}\pi^+\pi^-$ system, that are floated in the fit with Gaussian constraints.  The external values taken for the 
flavour-tagged fractions $K_i$ are those reported in Ref.~\cite{NEWWINGS}, and arise from a study of the results of  amplitude models developed by the BaBar and Belle collaborations~\cite{BELLE_2010,BABAR_2010,BABAR_2005,BABAR_2008}.  The values of the strong-phase parameters $c_i$ and $s_i$ come from quantum-correlated measurements performed by CLEO~\cite{CLEOKSPIPI}.

The best fit values and the correlations for the  parameters of interest are given in Tables~\ref{tab:cleocresults} and \ref{tab:correl_cleoc}, respectively.  
Note that the precision on $r_D^{K3\pi}$ and $r_D^{K\pi\pi^0}$ is limited by the knowledge of the ratios of the branching fractions, rather than the CLEO-c data.
Also shown are the previously reported results from Ref.~\cite{NEWWINGS}.   A significant change is observed in the value of the $D \to \KPIPIPI$ coherence factor, which is about 1$\sigma$ higher than previously.   This  shift is mainly driven by the change in the like-sign observables. 
A large change is also found in the central value of $\delta_D^{K3\pi}$, which is the result of the correction implemented in Eq.~\ref{eq:kspipiyield}.
As expected there is very little change in the $D \to \KPIPIZ$ results, as here the only input observable that has evolved is $\rho^{K3\pi}_{K\pi\pi^0, LS}$.   The reduced $\chi^2$ of the fit is 29.5/33, to be compared with 44.4/33 for the previous analysis.  Hence the compatibility of the input observables has improved.

\begin{table}[htb]
\caption{Results from the fit to the CLEO-c observables. The uncertainties are the combination of the statistical and systematic uncertainties. The results from the previous update~\cite{NEWWINGS}, where available, are also shown for comparison.} \label{tab:cleocresults}
\begin{center}
\begin{tabular}{lcc} \hline\hline
Parameter & Fitted value &  Previous result \\ \hline
\noalign{\vskip 0.075cm}
$R_{K3\pi}$                       &  $0.53^{+0.18}_{-0.21}$  & $0.32^{+0.20}_{-0.28}$\\
$\delta_D^{K3\pi}$           &  $(125^{+22}_{-14})^{\circ}$  &  $(255^{+21}_{-78})^{\circ}$ \\ 
$r_D^{K3\pi}$ & $(5.50\pm 0.12)\times 10^{-2}$ & -- \\
$R_{K\pi\pi^{0}}$             &  $0.82\pm 0.06$          &   $0.82\pm 0.07$  \\ 
$\delta_D^{K\pi\pi^{0}}$ &  $(199^{+13}_{-14})^{\circ}$     &  $(164^{+20}_{-14})^{\circ}$\\ 
$r_D^{K\pi\pi^0}$ & $(4.48\pm 0.12)\times 10^{-2}$ &  -- \\ \hline\hline
\end{tabular}
\end{center}
\end{table}

\begin{table}[htb]
\caption{Correlation coefficients between the parameters in the fit to the  CLEO-c observables.}\label{tab:correl_cleoc} 
\begin{center}
\begin{tabular}{l|rrrrrr}
                                 & $R_{K3\pi}$ & $\delta_D^{K3\pi}$  & $r_D^{K3\pi}$ & $R_{K\pi\pi^{0}}$ & $\delta_D^{K\pi\pi^0}$ & $r_D^{K\pi\pi^0}$ \\ \hline
$R_{K3\pi}$                   & $1.00$ &           $-0.47$ &           $-0.45$ & $\phantom{-}0.06$ &           $-0.01$ & $\phantom{-}0.08$ \\ 
$\delta_D^{K3\pi}$       &       & $\phantom{-}1.00$ & $\phantom{-}0.04$ & $\phantom{-}0.02$ & $\phantom{-}0.24$ & $\phantom{-}0.06$ \\
$r_D^{K3\pi}$               &     &               & $\phantom{-}1.00$ &           $-0.03$ &           $-0.06$ &           $-0.04$ \\ 
$R_{K\pi\pi^{0}}$         &     &               &               & $\phantom{-}1.00$ & $\phantom{-}0.25$ &           $-0.03$ \\ 
$\delta_D^{K\pi\pi^0}$ &    &               &               &               & $\phantom{-}1.00$ &           $-0.04$ \\ 
$r_D^{K\pi\pi^0}$         &    &               &               &               &               & $\phantom{-}1.00$ \\
\end{tabular}
\end{center}
\end{table}

\begin{figure}
\begin{center}
\begin{tabular}{cc}
\includegraphics[width=0.45\columnwidth]{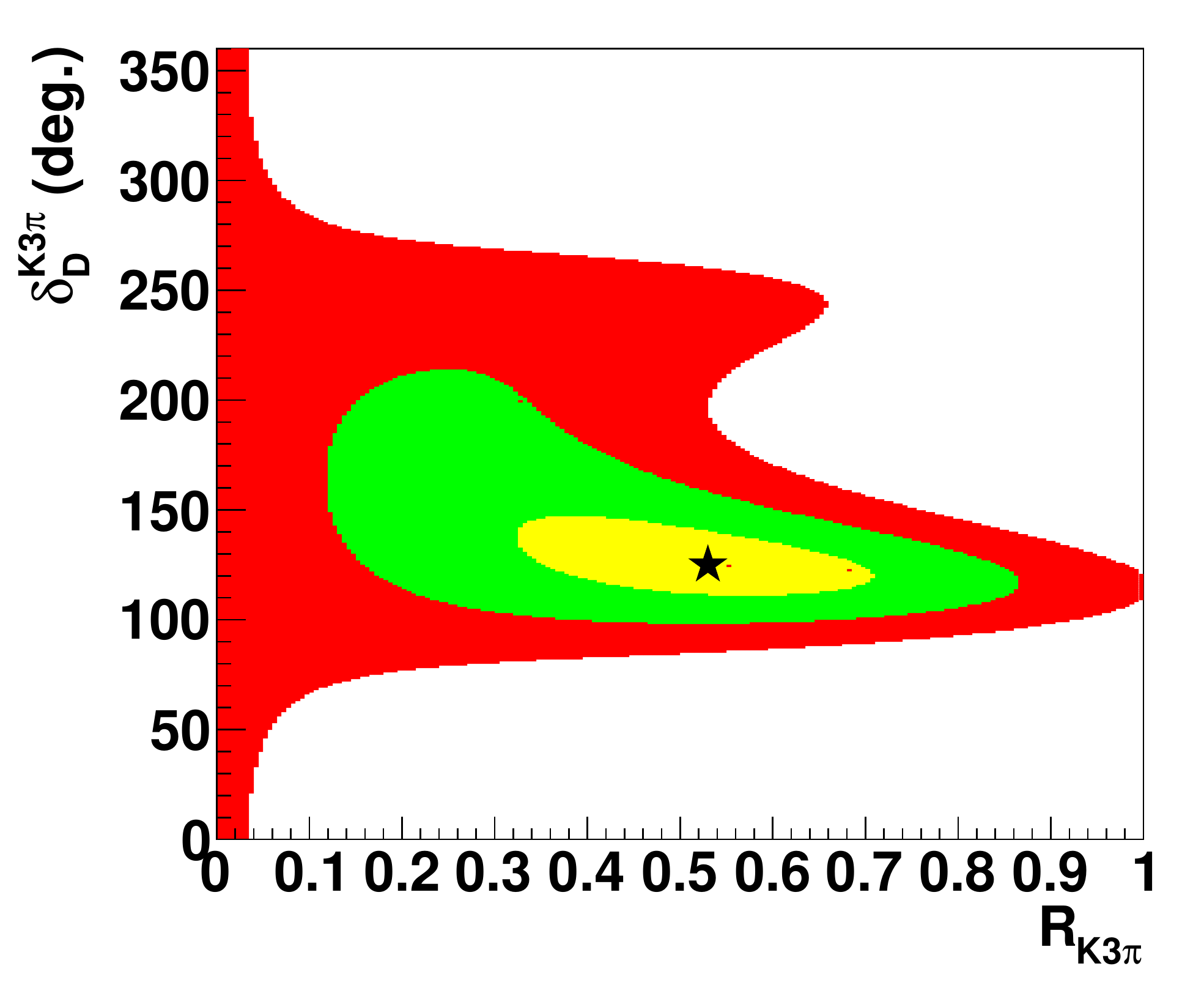}  &
\includegraphics[width=0.45\columnwidth]{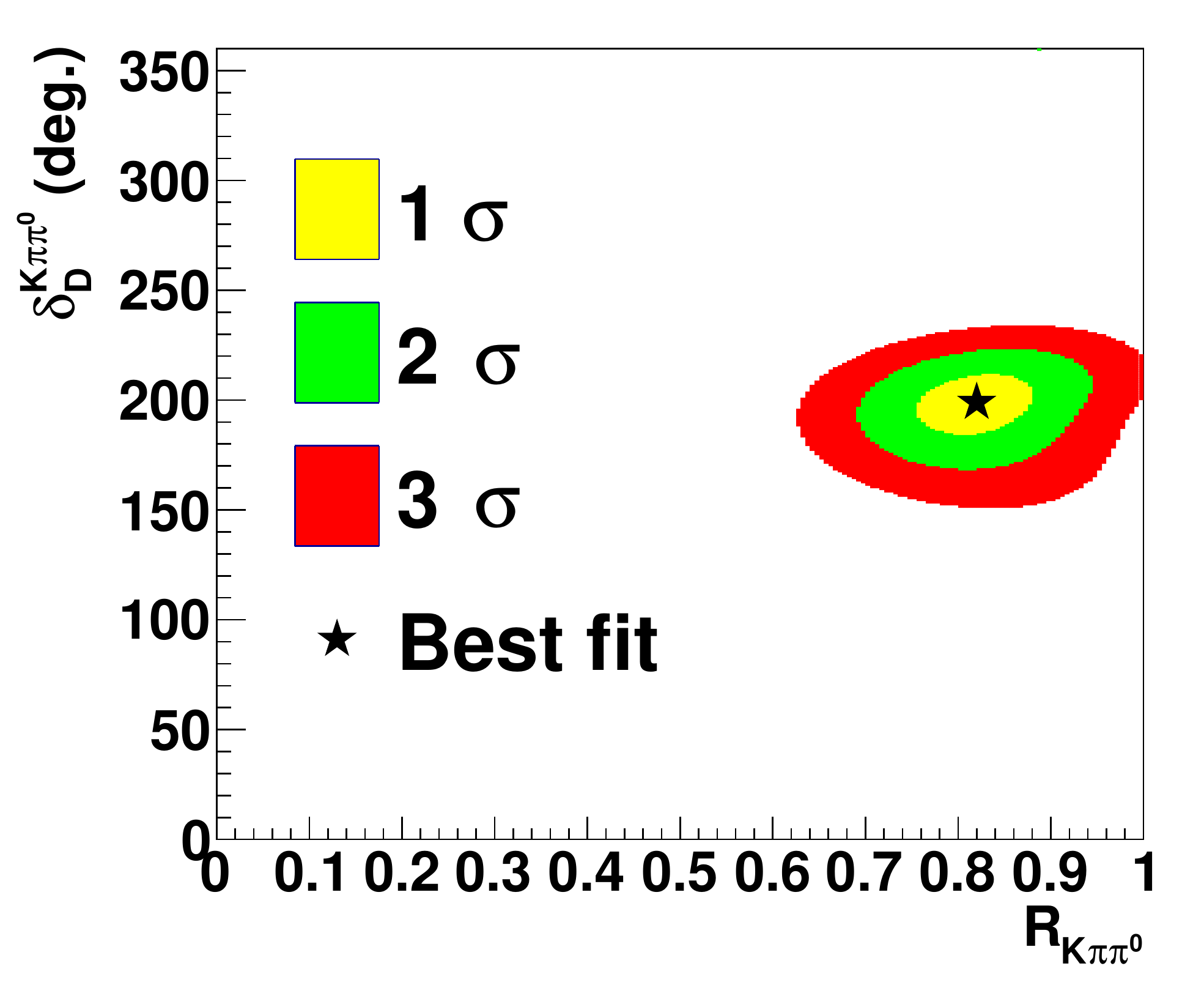}  \\
\end{tabular}
\caption{ Scans of $\Delta\chi^2$ for the fit to the updated CLEO-c observables in the (left) $(R_{K3\pi},\delta_{D}^{K3\pi})$ and (right) $(R_{K\pi\pi^{0}},\delta_{D}^{K\pi\pi^{0}})$ parameter space, showing the $\Delta \chi^2 =1, 4$ and $9$ intervals. }\label{fig:scans}
\end{center}
\end{figure}

Scans of the $(R_{K3\pi},\delta_{D}^{K3\pi})$ and $(R_{K\pi\pi^{0}},\delta_{D}^{K\pi\pi^{0}})$  parameter space are shown in Fig.~\ref{fig:scans}. %$\Delta\chi^2$ is used to determine the one, two and three standard deviation confidence intervals within the parameter space. 
In making these plots the values of $R$ and $\delta_D$ are fixed, while all other parameters 
are refitted to obtain $\Delta \chi^2$, the change in $\chi^2$ with respect to the lowest value found.

\section{Constraints on the coherence factor and average strong-phase difference from LHCb data}
\label{sec:combolhcb}

The LHCb collaboration has performed a study of the time-dependence of the ratio between $D^0 \to K^+\pi^-\pi^+\pi^-$ and  $D^0 \to K^-\pi^+\pi^+\pi^-$ decay rates~\cite{LHCBMIX}.  Several sets of results are reported, including those given in Table~\ref{tab:lhcbinputs} from a fit for $r_D^{K3\pi}$ and the parameters $a$ and $b$, assuming the functional form given in Eq.~\ref{eq:mixing}.

\begin{table}[tbh]
\caption{Results from the `unconstrained' time-dependent $D^0 \to \KPIPIPI$ analysis of LHCb~\cite{LHCBMIX}.}
\label{tab:lhcbinputs}
\begin{center}
\begin{tabular}{cc} \hline\hline
Parameter  &  Result \\ \hline
\noalign{\vskip 0.075cm}
$r_D^{K3\pi}$ &  $(5.67 \pm 0.12) \times 10^{-2}$ \\
$a$  &  $(0.3 \pm 1.8 ) \times 10^{-3}$ \\
$b$  &  $(4.8 \pm 1.8 ) \times 10^{-5}$ \\ \hline\hline
\end{tabular} 
\end{center}
\end{table}

\begin{figure}
\begin{center}
\includegraphics[width=0.65\columnwidth]{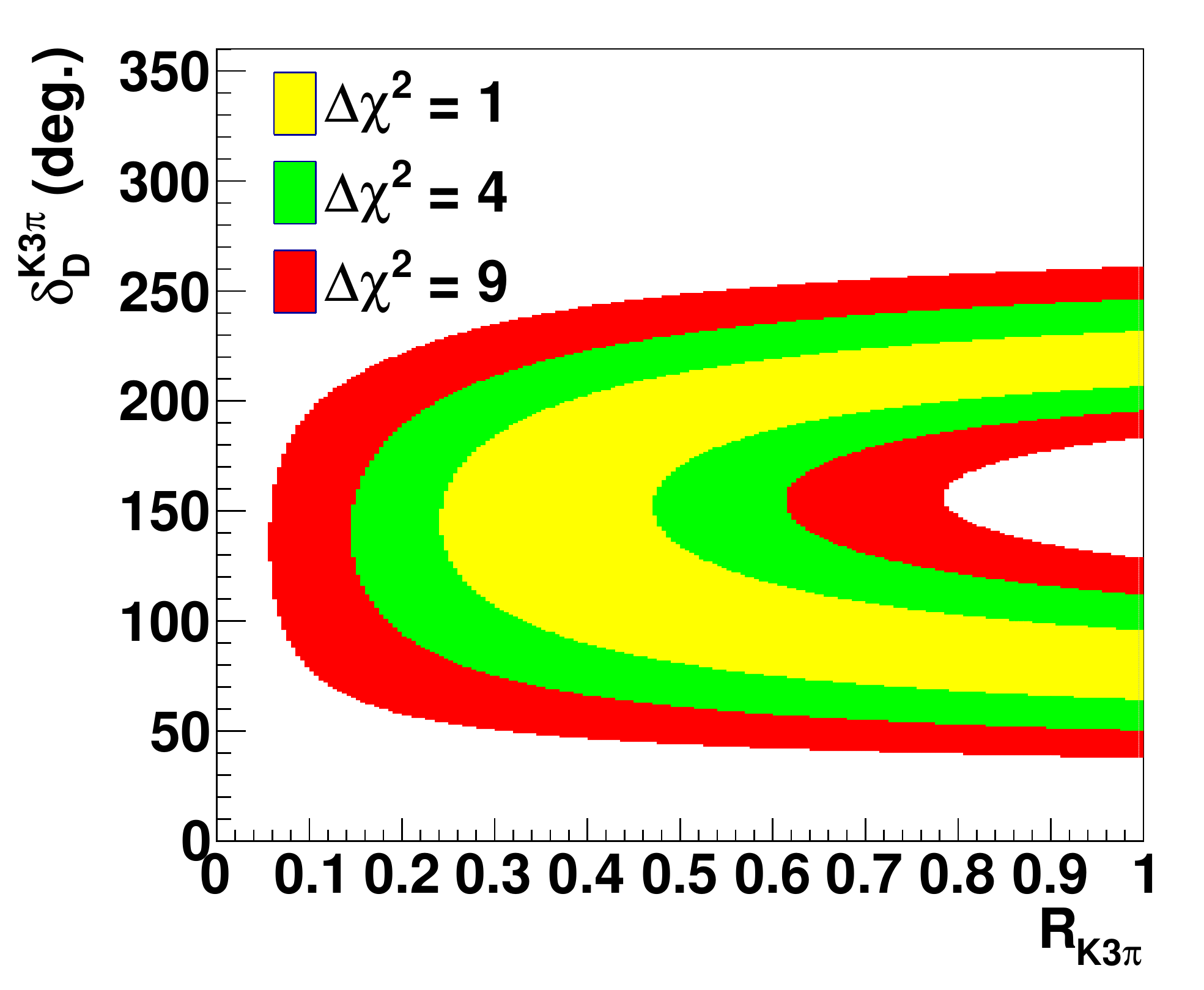}  
\caption{ Scan of $\Delta \chi^2$ in $(R_{K3\pi},\delta_{D}^{K3\pi})$  parameter space corresponding to the results of the time-dependent  $D^0 \to \KPIPIPI$ analysis of LHCb. }\label{fig:lhcbscan}
\end{center}
\end{figure}

Figure~\ref{fig:lhcbscan} shows a $\Delta \chi^2$  scan of the $(R_{K3\pi},\delta_{D}^{K3\pi})$ parameter space obtained from the LHCb results, and imposing Gaussian constraints on the mixing parameters $x$ and $y$ according to the measured values in Table~\ref{tab:inputs}.   The shape of the contours is significantly  different to those obtained from the fit to the CLEO-c observables, therefore motivating a combined fit of both sets of measurements.

\section{Combined fit}
\label{sec:combofit}

\begin{figure}
\begin{center}
\begin{tabular}{cc}
\includegraphics[width=0.45\columnwidth]{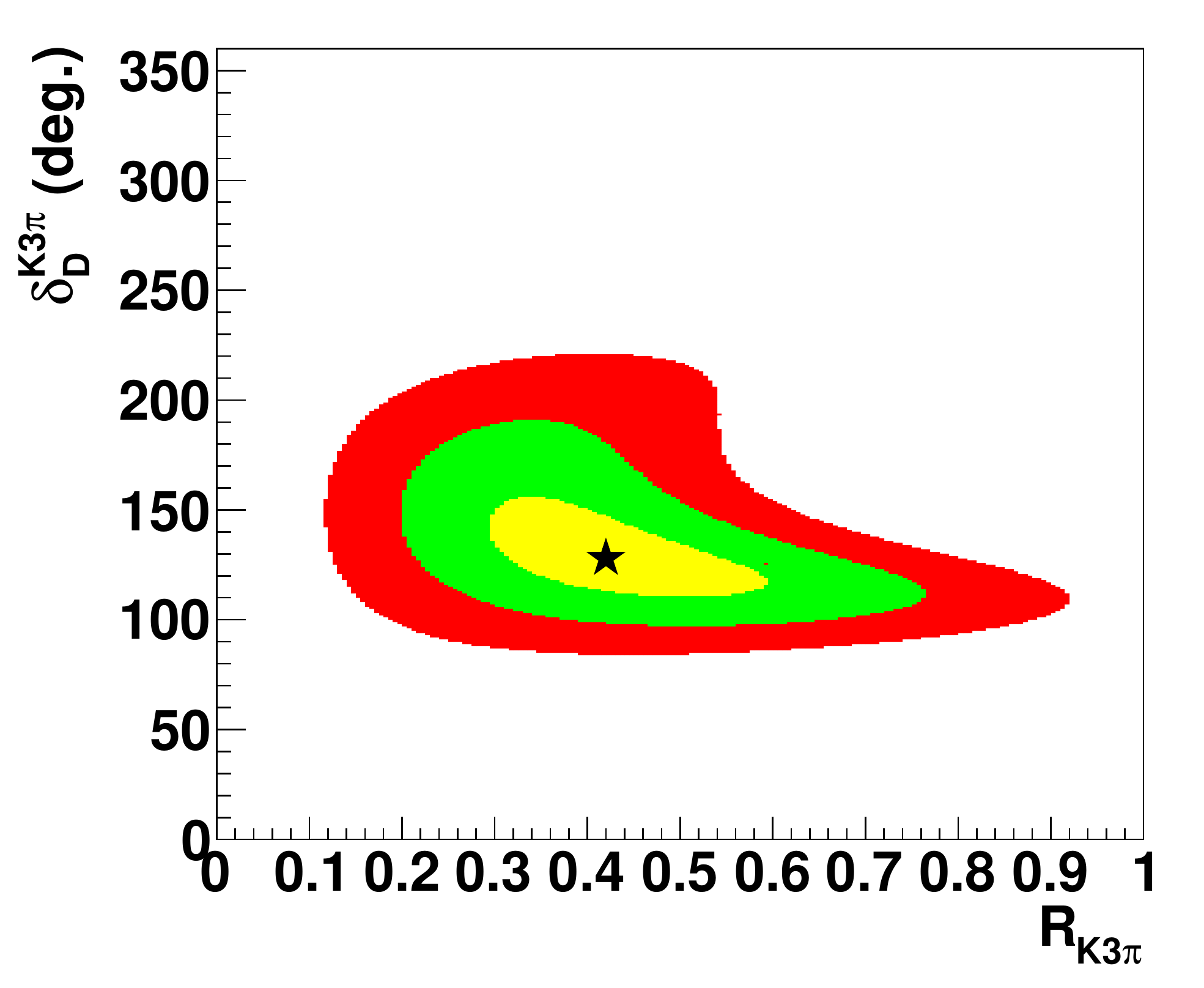}  &
\includegraphics[width=0.45\columnwidth]{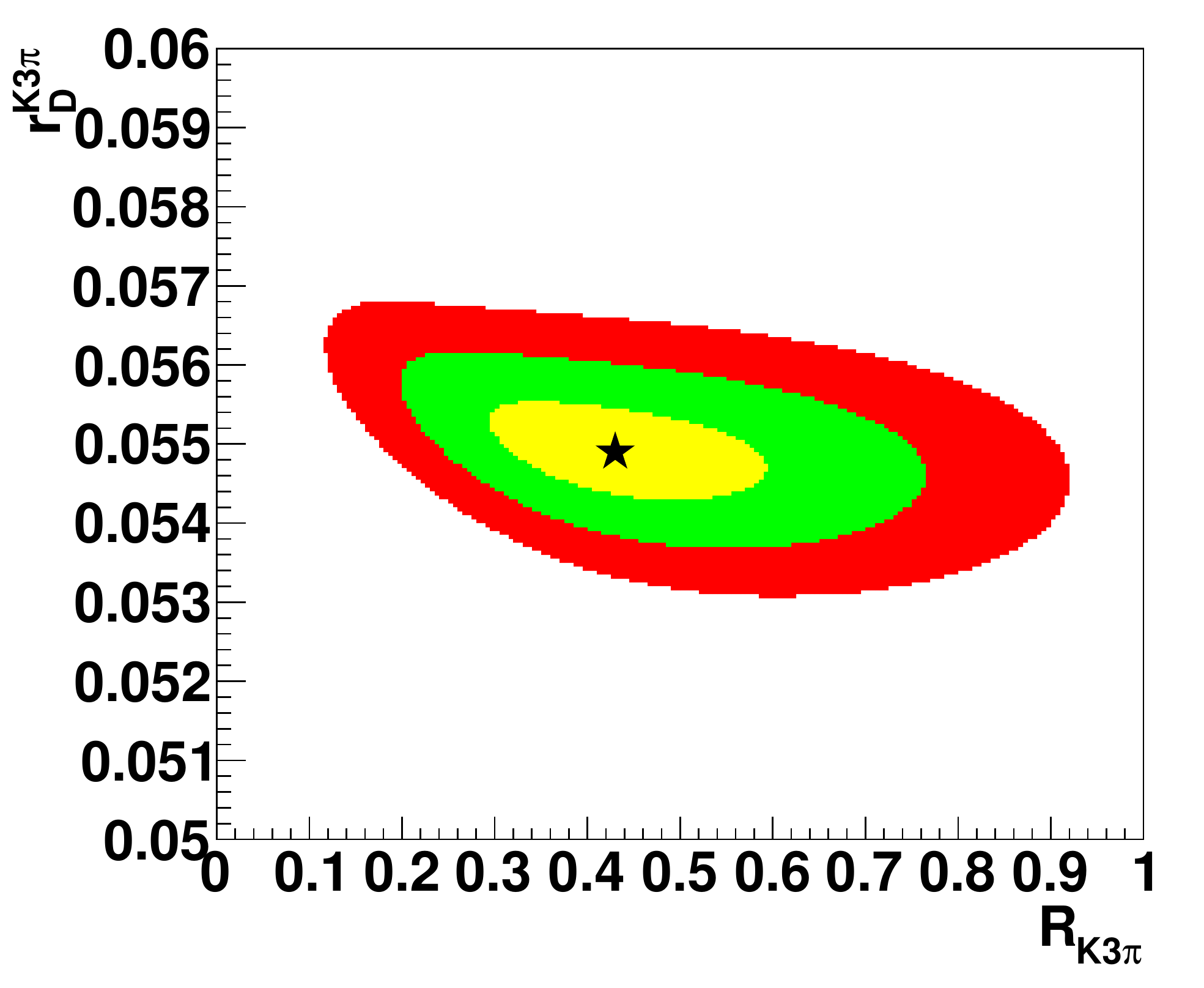}  \\
\includegraphics[width=0.45\columnwidth]{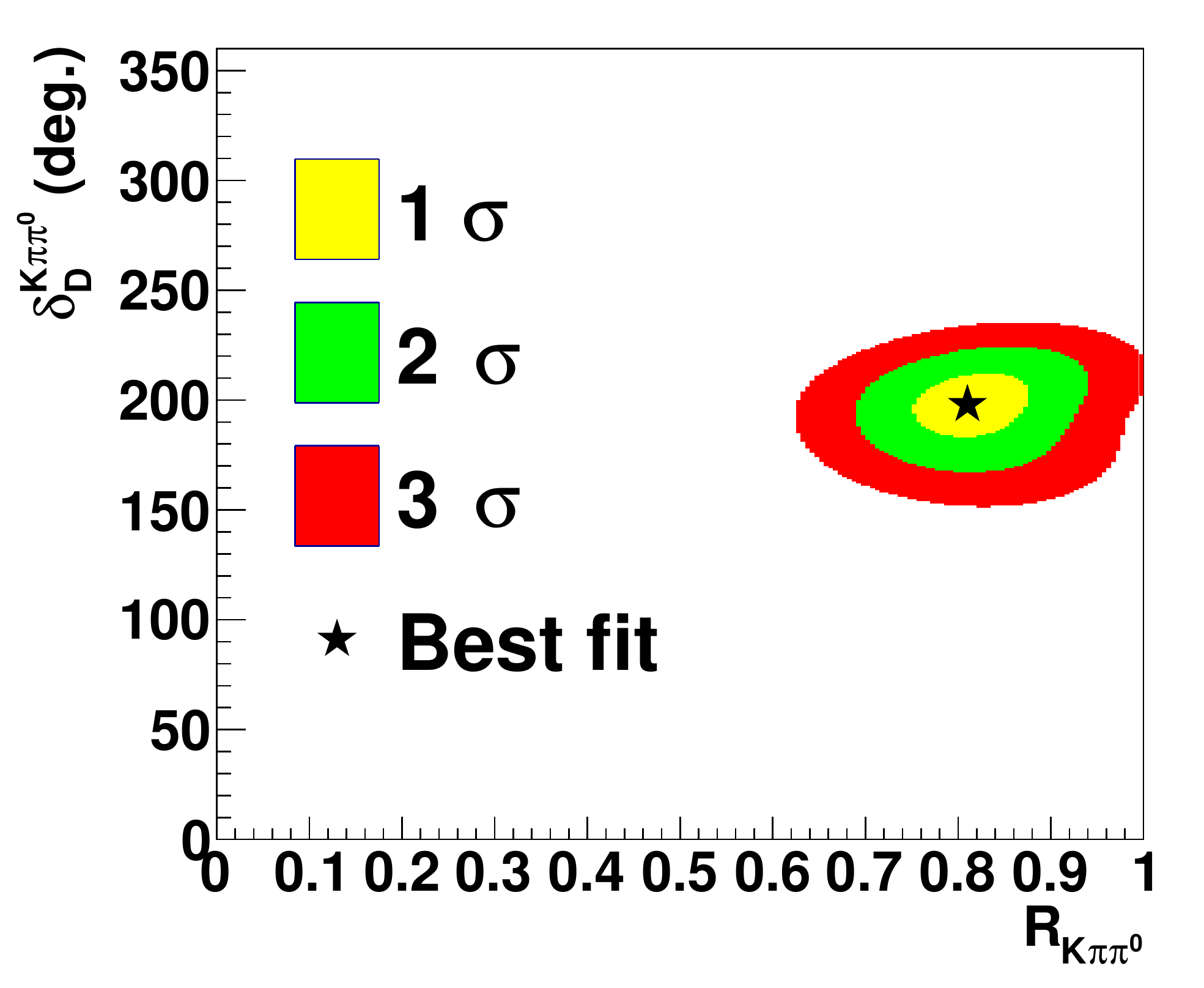}  &
\includegraphics[width=0.45\columnwidth]{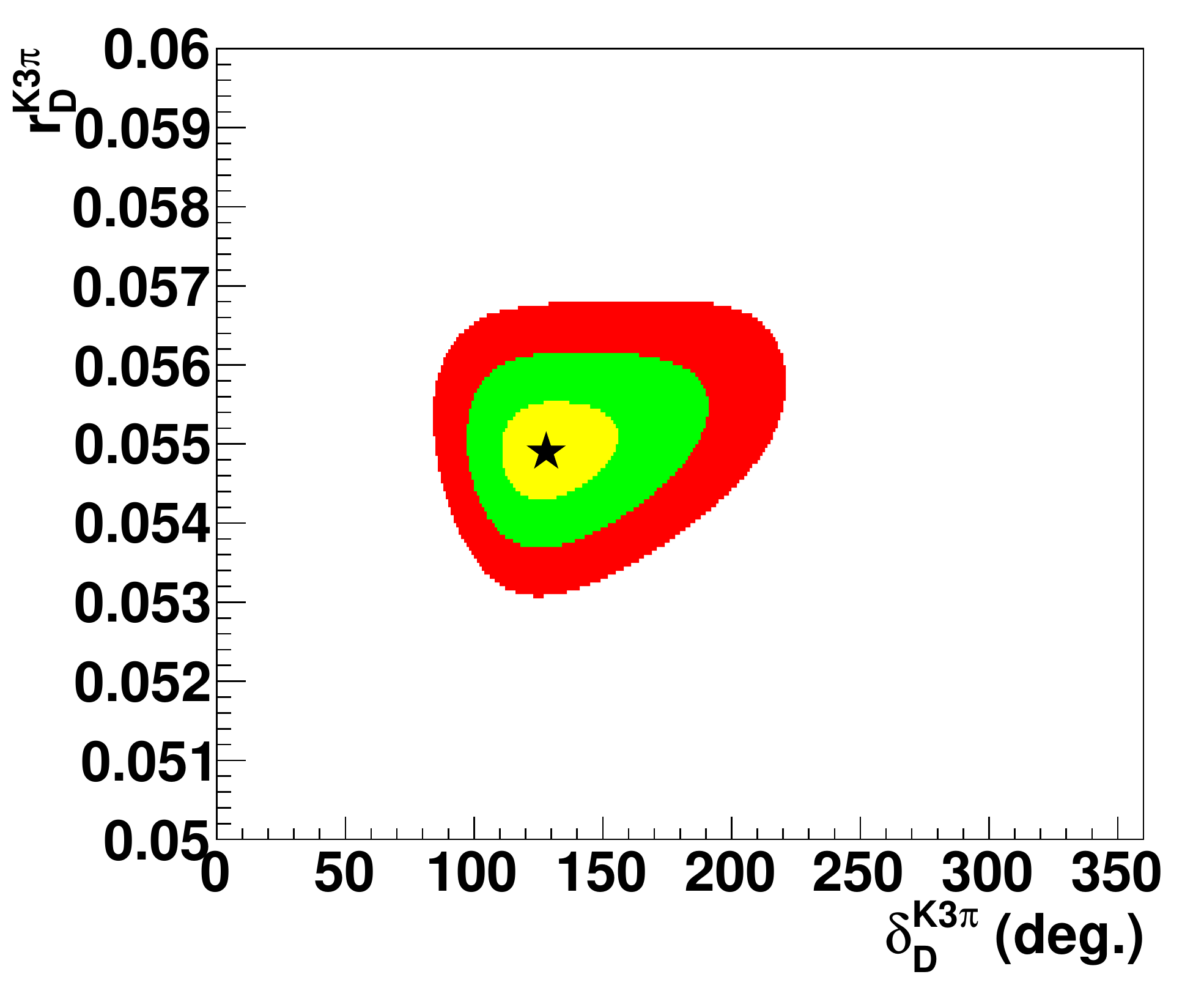} \\
\end{tabular}
\caption{ Scans of $\Delta\chi^2$ for the combined fit to the updated CLEO-c and LHCb observables in the (clockwise from top left) $(R_{K3\pi},\delta_{D}^{K3\pi})$, $(R_{K3\pi}, r_D^{K3\pi})$, $(\delta_D^{K3\pi},  r_D^{K3\pi})$ and $(R_{K\pi\pi^0},\delta_{D}^{K\pi\pi^0})$ parameter space.}\label{fig:scans_final}
\end{center}
\end{figure}

The fit described in Sect.~\ref{sec:cleofit} is repeated with the LHCb $D^0\bar{D}^0$-mixing results (reported in Table~\ref{tab:lhcbinputs})  included as additional input measurements.   The best fit values for the hadronic parameters, and the associated correlations, are presented in Tables~\ref{tab:finalresults} and~\ref{tab:correl_final}, respectively.    The reduced $\chi^2$ of the fit is 33.5/36.   Figure~\ref{fig:scans_final} shows the three possible sets of two-dimensional scans  in the  $D \to \KPIPIPI$ hadronic-parameter space; also shown is a scan of  $(R_{K\pi\pi^0},\delta_{D}^{K\pi\pi^0})$.    The inclusion of the LHCb observables improves the precision of the $D \to \KPIPIPI$ coherence factor, but lowers the central value with respect to that returned by the CLEO-c fit.  In this region the 1$\sigma$ bound on $\delta_D^{K3\pi}$ is weaker, although the results for this parameter become significantly  more Gaussian in behaviour.
The reduction in the uncertainty on $r_D^{K3\pi}$ is largely driven by the correlation with the mixing parameters $x$ and $y$, which are constrained through external measurements in the fit.  As expected there are only minor changes in the $D \to \KPIPIZ$ results compared to those obtained from the fit to the CLEO-c observables alone.

\begin{table}[ht]
\caption{Results from the combined fit to the updated CLEO-c and LHCb observables. The uncertainties are the combination of the statistical and systematic uncertainties. } \label{tab:finalresults}
\begin{center}
\begin{tabular}{lccc} \hline\hline
Parameter &  Fitted value \\ \hline
\noalign{\vskip 0.075cm}
$R_{K3\pi}$                       &   $0.43^{+0.17}_{-0.13}$  \\
$\delta_D^{K3\pi}$           &   $(128^{+28}_{-17})^{\circ}$  \\ 
$r_D^{K3\pi}$ &   $(5.49\pm 0.06)\times 10^{-2}$ \\
$R_{K\pi\pi^{0}}$             &    $0.81 \pm 0.06$\\ 
$\delta_D^{K\pi\pi^{0}}$ &    $(198^{+14}_{-15})^{\circ}$ \\ 
$r_D^{K\pi\pi^0}$ &   $(4.47 \pm 0.12)\times 10^{-2}$  \\ \hline\hline
\end{tabular}
\end{center}
\end{table}

\begin{table}[htb]
\caption{Correlation coefficients between the parameters from the combined fit to the updated CLEO-c and LHCb observables.}\label{tab:correl_final} 
\begin{center}
\begin{tabular}{l|rrrrrr}
                                 & $R_{K3\pi}$ & $\delta_D^{K3\pi}$  & $r_D^{K3\pi}$ & $R_{K\pi\pi^{0}}$ & $\delta_D^{K\pi\pi^0}$ & $r_D^{K\pi\pi^0}$ \\ \hline
$R_{K3\pi}$                   &  $1.00$ &           $-0.67$ &           $-0.48$ & $\phantom{-}0.03$ &           $-0.05$ &           $-0.04$ \\
$\delta_D^{K3\pi}$       &   & $\phantom{-}1.00$ & $\phantom{-}0.12$ & $\phantom{-}0.02$ & $\phantom{-}0.15$ & $\phantom{-}0.08$ \\ 
$r_D^{K3\pi}$               &     &               & $\phantom{-}1.00$ &           $-0.04$ &           $-0.02$ &           $-0.03$ \\  
$R_{K\pi\pi^{0}}$         &     &              &               & $\phantom{-}1.00$ & $\phantom{-}0.23$ &           $-0.04$ \\ 
$\delta_D^{K\pi\pi^0}$ &    &               &              &              & $\phantom{-}1.00$ &           $-0.03$ \\  
$r_D^{K\pi\pi^0}$         &    &              &               &              &              & $\phantom{-}1.00$ \\
\end{tabular}
\end{center}
\end{table}

\section{Conclusions}

A re-analysis of the CLEO-c $\psi(3770)$ data set has yielded an updated set of observables sensitive to the hadronic parameters of the decay $D \to \KPIPIPI$, some of which are significantly different to those reported previously~\cite{WINGS,NEWWINGS}.  These observables have been input to a combined fit, together with measurements from a recent LHCb $D^0 \bar{D}^0$ mixing analysis~\cite{LHCBMIX}.    Results are obtained for $R_{K3\pi}$ and   $r_D^{K3\pi}$ that are significantly more precise than those derived from the CLEO-c observables alone.  New values and constraints are also determined for the hadronic parameters of the decay $D \to \KPIPIZ$.   These results will be valuable for improving sensitivity to the unitarity triangle angle $\gamma$ with analyses exploiting $B^- \to D K^-$ decays.   The combined fit can be re-performed when future measurements of the $\psi(3770)$ observables become available from the BESIII collaboration, or when improved $D^0 \bar{D}^0$ mixing results are reported by either LHCb or Belle-II.

\section*{Acknowledgments}

This analysis was performed using CLEO-c data. The authors of this Letter (some of whom were members of CLEO) are grateful to the collaboration for the privilege of using these data. We also thank Roy Briere, David Cassel, Tim Gershon and Sheldon Stone for their careful reading of the draft manuscript and valuable suggestions.
We are grateful for support from 
the UK Science and Technology Facilities Council, the UK India and Education Research Initiative and the European Research Council under FP7.


\appendix

\section{Correlation matrix for  the observables measured with the CLEO-c data}
\label{sec:cleoccorrel}

The correlation matrix for the $D \to \KPIPIPI$ and $D \to \KPIPIZ$ inclusive observables is presented in Table~\ref{tab:correlfull}.  The $Y_i$ observables are all uncorrelated.

\begin{table}[thb]
\caption{Correlation matrix for the $D \to K^-\pi^+\pi^+\pi^-$ and $D \to K^-\pi^+\pi^0$  inclusive observables.}
\label{tab:correlfull}
\begin{center}
\begin{tabular}{l|ccccccc}
& $\Delta^{K3\pi}_{\mathit{CP}}$     & $\rho^{K3\pi}_{LS}$ &   $\rho^{K3\pi}_{K\pi, LS}$  & $\rho_{K\pi\pi^0,LS}^{K3\pi}$ & $\Delta^{K\pi\pi^0}_{\mathit{CP}}$    & $\rho^{K\pi\pi^0}_{LS}$  &   $\rho^{K\pi\pi^0}_{K\pi, LS}$ \\ \hline
$\Delta^{K3\pi}_{\mathit{CP}}$               &1.00   &   0.01  &   0.01     &  0.00  & 0.60  & 0.00   &  0.00 \\
 $\rho^{K3\pi}_{LS}$                 &           &    1.00 &  0.16  &  0.19  & 0.00  & 0.00   & 0.00   \\
$\rho^{K3\pi}_{K\pi, LS}$          &           &           &   1.00    &  0.20  & 0.00  & 0.00  & 0.00   \\
 $\rho_{K\pi\pi^0,LS}^{K3\pi}$ &           &            &                 &  1.00 & 0.03   &  0.02 &  0.00  \\
$\Delta^{K\pi\pi^0}_{\mathit{CP}}$        &           &            &                 &          & 1.00 &  0.03 &  0.01 \\
$\rho^{K\pi\pi^0}_{LS}$            &           &            &                 &           &         &  1.00    &  0.01 \\
$\rho^{K\pi\pi^0}_{K\pi, LS}$    &           &            &                &           &         &             &  1.00  \\

\end{tabular}
\end{center}
\end{table}




%% Section 1

\end{document}